\documentclass[aps,twocolumn,superscriptaddress,showpacs,floats]{revtex4}
\usepackage{amsmath}
\usepackage{amsfonts}
\usepackage{amssymb}
\usepackage{euscript}
\usepackage{bm}
\usepackage[dvips]{graphicx}

\begin{document}

\title{Kolmogorov and Kelvin-Wave Cascades of Superfluid Turbulence at $T=0$:\\ What is in Between ?}

\author{Evgeny Kozik}
 \affiliation{Department of Physics, University of
Massachusetts, Amherst, MA 01003}
\author{Boris Svistunov}
\affiliation{Department of Physics, University of Massachusetts,
Amherst, MA 01003} \affiliation{Russian Research Center
``Kurchatov Institute'', 123182 Moscow, Russia}

\begin{abstract}
As long as vorticity quantization remains irrelevant for the
long-wave physics, superfluid turbulence supports a regime
macroscopically identical to the Kolmogorov cascade of a normal
liquid. At high enough wavenumbers, the energy flux in the
wavelength space is carried by individual Kelvin-wave cascades on
separate vortex lines. We analyze the transformation of the
Kolmogorov cascade into the Kelvin-wave cascade, revealing a chain
of three distinct intermediate cascades, supported by
local-induction motion of the vortex lines, and distinguished by
specific reconnection mechanisms. The most prominent qualitative
feature predicted is unavoidable production of vortex rings of the
size of the order of inter-vortex distance.

\end{abstract}


%
%
%

\maketitle

Nowadays, superfluid turbulence \cite{Donnelly,Cambridge}---a
structured or non-structured tangle of quantized vortex lines---is
attracting much attention \cite{Vinen06}, stimulated, in
particular, by advances in experimental techniques allowing
studies of different turbulent regimes in diverse superfluid
systems, such as $^4$He \cite{Cambridge, Lathrop}, $^3$He-B
\cite{3He_Nature, Pickett}, and Bose-Einstein condensates of
ultacold atoms \cite{Cambridge,BEC}. In superfluids at $T=0$,
vorticity can only exist in the form of topological
defects---vortex lines of microscopic thickness, the circulation
of velocity around which being equal to the liquid-specific
quantum $\kappa$. Speaking generally, the dynamical mechanisms
governing superfluid turbulence are fundamentally different from
those of classical turbulence (see, e.g., recent review
\cite{Vinen06} and references therein).

A new wave of interest in dynamics of superfluid turbulence came
with the experiment by Maurer and Tabeling \cite{Maurer_Tabeling},
who observed that superfluid turbulence in $^4$He  formed by
counter-rotating discs is indistinguishable from classical
turbulence at large length scales, in particular, exhibiting the
classical Kolmogorov cascade. Shortly, the same effect was found in
superfluid turbulence generated by a towed grid \cite{Skrbek}. In
the experiments \cite{Maurer_Tabeling, Skrbek}, the fraction of
normal component is considerable making analysis of vortex tangle
dynamics and structure significantly complicated
\cite{Vinen2000,Vinen_Niemela}. (Considerations regarding possible
energy spectra in this case are presented in
Ref.~\cite{En_spectra}.) However, the similarity between classical
and superfluid turbulence exists even at practically zero
temperature, which was first observed in numerical simulations
\cite{Nore, Tsubota_2002, Barenghi_2002}, and, just recently, for
the first time confirmed by measurements in $^3$He-B \cite{Pickett}.

\begin{figure}[htb]
\includegraphics[width = 0.95\columnwidth,keepaspectratio=true]{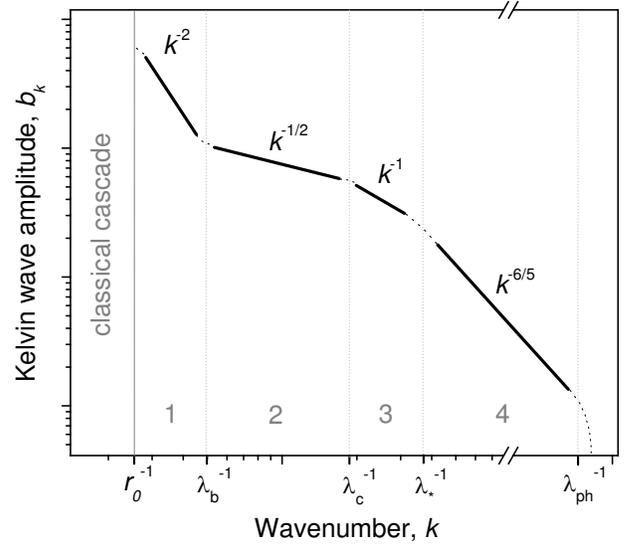}
\caption{Spectrum of Kelvin waves in the quantized regime. The
inertial range consists of a chain of cascades driven by different
mechanisms: (1) reconnections of vortex-line bundles, (2)
reconnections between nearest-neighbor vortex lines in a bundle,
(3) self-reconnections on single vortex lines, (4) non-linear
dynamics of single vortex lines without reconnections. The regimes
(3) and (4) are familiar in the context of non-structured vortex
tangle decay \cite{Sv_95, SvK_2004}.}
\label{f1}
\end{figure}

By the nature of a cascade regime, implying that the kinetic times
get progressively shorter down the hierarchy of length scales,
instantaneous structure of turbulence follows the evolution at the
largest length scales (typically of order of system size), where the
energy flux (per unit mass) $\varepsilon$ is formed. At very low
temperatures, due to the absence of frictional dissipation, the flux
$\varepsilon$ must be carried down to scales significantly smaller
than the (related to $\varepsilon$) typical separation between the
vortex lines $l_0$. At small enough length scales, the energy flux
is carried by pure Kelvin-wave cascades on separate vortex lines
\cite{SvK_2004, Vinen_2003}, the cutoff being due to sound radiation
\cite{Vinen2000, Vinen2001, SvK_vortex_phonon}.

The fact that superfluid turbulence at large compared to $l_0$
length scales may be consistent with the classical Kolmogorov law is
not surprising (a formal proof is mentioned below). It is well known
\cite{Donnelly} that macroscopic velocity profile of a rapidly
rotated superfluid mimics solid-body rotation, which is accomplished
by formation of a dense array of vortex lines aligned along the
rotation axis. By the same mechanism, ``stirring'' a superfluid one
can produce vorticity in the \textit{course-grained} up to length
scales larger than $l_0$ superfluid velocity field,
indistinguishable from that of a normal fluid, the underlying vortex
tangle being organized in polarized ``bundles'' of vortex lines.
What turns out to be a puzzle \cite{Bottleneck}, however, is how the
vortex tangle looks like when one zooms in down to scales of order
of the interline separation $l_0$, where the vorticity is
essentially discrete.

In this Letter, we analyze the structure of turbulence at all
length scales tracing the transformation of the \textit{classical}
regime, described by the Kolmogorov law at large length scales,
into the \textit{quantized} regime, in which the discreteness of
vortex lines is important, in the fundamental case of zero
temperature. The analysis relies on the large parameter
\begin{equation}
\Lambda\,  =\,  \ln(l_0/a_0) \gg 1, \label{Lambda}
\end{equation}
where $a_0$ is the vortex core radius. In realistic $^4$He
experiments, $\Lambda \sim 15$. The attention to the problem of
linking the two regimes was drawn recently by L'vov \textit{et
al.} \cite{Bottleneck}, who realized that it is impossible to
directly cross over from Kolmogorov regime to the pure Kelvin-wave
cascade, and put forward the idea of a bottleneck, with specific
dynamical implications. The Achilles' heel  of the treatment of
Ref.~\cite{Bottleneck} is taking it for granted \cite{Vinen2000,
Vinen_Niemela} that the coarse-grained  macroscopic description of
quantized vorticity remains valid down to the scale of $l_0$.

We show that the locally-induced motion of the vortex lines
radically changes dynamical picture already at the scale
\begin{equation}
r_0 \sim \Lambda^{1/2} l_0\, ,\label{r_0-l_0}
\end{equation}
with the interline separation related to the energy flux by
\begin{equation}
l_0 \sim ( \Lambda \kappa^3/\varepsilon)^{1/4} \; . \label{l_0}
\end{equation}
In the range of wavelength $r_0 > \lambda >
\lambda_*$,
\begin{equation}
\lambda_*=l_0/\Lambda^{1/2}, \label{lambda_*}
\end{equation}
there takes place a chain of three cascade regimes, in which the
energy flux $\varepsilon$ is carried by locally-induced motion
combined with vortex line reconnections. The three regimes are
distinguished by their specific types of reconnections: (i)
reconnections of vortex-line bundles, (ii) reconnections between
nearest-neighbor vortex lines, (iii) self-reconnections on single
vortex lines---the mechanism introduced earlier by one of us
\cite{Sv_95} in the context of the decay of non-structured
superfluid turbulence. The existence of the regime (iii) means
unavoidable production of vortex rings of the typical size
$\lambda_*$ at a
rate immediately following from (\ref{l_0}) by conservation of
energy. Namely, $\sim \kappa \Lambda^{1/2}/l_0^2$ rings are emitted per unit
time in the characteristic volume of $l_0^3$. Remarkably, this
rate is $\sim \Lambda^{3/2}$ times smaller than the rate of vortex ring
production characteristic of non-structured superfluid turbulence
\cite{Sv_95, SvK_2004}.

For realistic values of $\Lambda$, sharp distinction between the
three sub-regimes is likely to be lost, although characteristic
features of strong turbulence, such as generation of a spectrum of
vortex rings by the mechanism (iii), might manifest themselves. At
the wavelength scale $\lambda_*$, self-reconnections cease and the
weak-turbulent regime sets in, with purely non-linear Kelvin-wave
cascade \cite{SvK_2004}. This regime covers a significant part of
the inertial range until eventually, at the scale
\cite{SvK_vortex_phonon}
\begin{equation}
\lambda_\mathrm{ph}= \left[\;\Lambda^{27}\; (\kappa/c)^{25}\;
l_0^{6}\;\right]^{1/31} \label{lambda_ph}
\end{equation}
($c$ is the sound velocity), the cascade is cut off due to the
radiation of sound by Kelvin waves. With $\kappa/c \sim a_0$ we
have $\lambda_\mathrm{ph} \lll \lambda_*$.

The structure of turbulence is summarized in Fig.~(\ref{f1}). We
emphasize that the notion of energy spectrum $E(k)$, where
$E(k)dk$ gives the energy per unit mass associated with variations
of fluid velocity over length scales $\sim k^{-1}$ in the interval
$dk$, is practically meaningful only in the classical regime. In
the quantized regime, the relevant degrees of freedom are waves on
vortex lines, while even a perfectly straight vortex line has a
nontrivial spectrum $E(k) \sim k$. On the experimental side,
recently introduced vortex line visualization technique in $^4$He
\cite{Lathrop} could provide the most direct probe for the
quantized regime.

{\it Analysis}. The key to understanding vortex dynamics at zero
temperature is given by Kelvin-Helmholtz's theorem, which states
that a vortex line element moves with local fluid velocity.
Mathematically, this is reflected in the Biot-Savart equation
\cite{Donnelly},
\begin{gather}
\dot{\textbf{s}} = \mathbf{v}(\mathbf{s})\, , ~~~
\mathbf{v}(\mathbf{r})={\kappa \over 4 \pi} \int (\textbf{s}_0 -
\textbf{r}) \times  {\rm d}\textbf{s}_0 / |\textbf{s}_0 -
\textbf{r}|^3 \; . \label{BS}
\end{gather}
Here $\mathbf{v}(\mathbf{r})$ is the superfluid velocity field,
$\textbf{s}$ is the time-evolving radius-vector of the vortex line
element, the dot denotes differentiation with respect to time, the
vector $\textbf{s}_0$ has the same physical meaning as $\textbf{s}$,
understood as an integration variable, and the integration is along
all the vortex lines. In the long-wave limit, when discreteness of
vortex lines is irrelevant, Eqs.~(\ref{BS}) can be coarse grained:
The second equation turns into Biot-Savart resolvent restoring the
field ${\bf v}$ from its curl, ${\bf w}$. The first equation
coarse-grains into Euler's equation for incompressible fluid,
\begin{equation}
\dot{w}_\alpha \; =\; -v_\beta \partial_\beta w_\alpha \, + \,
w_\beta \partial_\beta v_\alpha  \; . \label{Euler}
\end{equation}
Here the first term in the r.h.s. comes from translation of the
vortex array by the flow, the second term originates from bending of
the array by velocity gradient. This formally proves the equivalence
of (structured) superfluid and normal-ideal-incompressible-fluid
turbulence. For a non-structured tangle, coarse graining trivially
leads to $v\equiv0$.

For our purposes, it is instructive to formally decompose the
integral (\ref{BS}) into the self-induced part,
$\mathbf{v}^\mathrm{SI}(\mathbf{s})$, for which the integration is
restricted to the vortex line containing the element $\mathbf{s}$,
and the remaining contribution induced by all the other lines
$\mathbf{v}^\mathrm{I}(\mathbf{s})$,
\begin{equation}
\mathbf{v}(\mathbf{s})=\mathbf{v}^\mathrm{SI}(\mathbf{s})+\mathbf{v}^\mathrm{I}(\mathbf{s}).
\label{v-decomposition}
\end{equation}
Since  the velocities $\mathbf{v}^\mathrm{SI}$ and
$\mathbf{v}^\mathrm{I}$ define the r.h.s. of Eq.~(\ref{BS}), the
competition between them is crucial for the problem.

The leading contribution to $\mathbf{v}^\mathrm{SI}$ is given by the
local induction approximation (LIA), which reduces the integral over
the vortex line to its local differential characteristics,
\begin{equation}
\mathbf{v}^\mathrm{SI}(\textbf{s}) =  \Lambda_{R} \, {\kappa \over 4
\pi}  \, \textbf{s}' \times \textbf{s}'', \;\;\;\;\;  \Lambda_{R} =
\ln (R/a_0)\; , \label{LIA}
\end{equation}
where the prime denotes differentiation with respect to the arc
length and $R$ is the typical curvature radius. A necessary
condition of applicability of the LIA is $\Lambda_{R} \gg 1$.
Taking into account that $\Lambda_R$ is a very weak function of
$R$, we shall treat it as a constant of typical value $\Lambda_{R}
\sim \Lambda$. Note, however, that despite the fact that the
condition (\ref{Lambda}) is typically well satisfied, using the
LIA is not always appropriate.---Being an integrable model, LIA
does not capture reconnection-free (purely non-linear) kinetics of
Kelvin waves \cite{SvK_2004}.

To determine the crossover scale $r_0$, consider the structure of
the vortex tangle in the classical regime. By the definition of
$r_0$, at length scales $r \gg r_0$ turbulence mimics classical
vorticity by taking on the form of a dense coherently moving array
of vortex lines bent at curvature radius of order $r$. Velocity
field of this configuration obeys the Kolmogorov law
\begin{equation}
v_r \sim (\varepsilon r)^{1/3}, ~~~~~~~~~ r \gg r_0\, .
\label{Kolmogorov}
\end{equation}
Here and below the subscript $r$ means typical variation of a
field over distance $\sim r$. On the other hand, the value of
$v_r$ is fixed by the quantization of velocity circulation around
a contour of radius $r$, namely $v_r r \sim \kappa n_r r^2$, where
$n_r$ is the areal density of vortex lines responsible for
vorticity at the scale $r$. Note, that scale invariance requires
that on top of vorticity at the scale $r$ there be a fine
structure of vortex bundles of smaller sizes, so that,
mathematically, $n_r r^2$ is the difference between huge numbers
of vortex lines crossing the area of the contour $r$ in opposite
directions. The quantity $n_r$ is related to the flux by
\begin{equation}
n_r \sim \left[ \frac{\varepsilon}{\, \kappa^3 \; r^2}
\right]^{1/3}\, , ~~~~~~ r \gg r_0\, . \label{n_r}
\end{equation}
The underlying dynamics of a single vortex line in the bundle is
governed by $v^\mathrm{I}_r$ and $v^\mathrm{SI}_r$. While, by its
definition, $v^{I}_r \sim v_r$, which is given by
Eq.~(\ref{Kolmogorov}), the self-induced part is determined by the
curvature radius $r$ of the vortex line according to
Eq.~(\ref{LIA}),
\begin{equation}
v^\mathrm{SI}_r \sim \Lambda \frac{\kappa }{r} . \label{v-SI_r}
\end{equation}
At length scales where $v^\mathrm{I}_r \gg v^\mathrm{SI}_r$, the
vortex lines in the bundle move coherently with the same velocity
$\sim v^\mathrm{I}_r$. However, at the scale $r_0 \sim (\Lambda^3
\kappa^3/\varepsilon)^{1/4}$, the self-induced motion of the vortex
line becomes comparable to the collective motion, $v^\mathrm{SI}_r
\sim v^\mathrm{I}_r $. At this scale, individual vortex lines start
to behave independently from each other and thus $r_0$ gives the
lower cutoff of the inertial region of the Kolmogorov spectrum
(\ref{Kolmogorov}).

Since $r_0$ is the size of the smallest classical eddies, the
areal density of vortex lines at this scale is given by the
typical interline separation, $n_{r_0} \sim 1/l_0^2$. With
Eq.~(\ref{n_r}), we arrive at (\ref{r_0-l_0})-(\ref{l_0}).

At the scale $r_0$, turbulence consists of randomly oriented
vortex line bundles of size $r_0$, left by the classical regime.
The typical number of vortices in the bundle is given by $n_{r_0}
r_0^2 \sim \Lambda$. The length $r_0$ plays the role of a
correlation radius in the sense that relative orientation of two
vortex lines becomes uncorrelated only if they are a distance
$\gtrsim r_0$ apart. On the other hand, the crossover to the
quantized regime means that each line starts moving according to
its geometric shape, as described by Eq.~(\ref{LIA}). Therefore,
reconnections, at least between separate bundles, are inevitable
and, as we show below, capable of sustaining the flux
$\varepsilon$.

Reconnections play the leading role at $r_0 \gtrsim \lambda
\gtrsim \lambda_*$. Although this region is relatively narrow as compared to the whole Kelvin-wave inertial range, it is significantly large in the absolute units. Before going into the details of the
reconnection-assisted regimes, we describe the remaining and
dominant region of the cascade. As was shown by the authors
\cite{SvK_2004}, at a sufficiently small wavelength, a strongly
turbulent cascade of Kelvin waves is replaced by a purely
non-linear cascade, in which the reconnections are exponentially
suppressed. The spectrum of Kelvin-wave amplitudes $b_k$, $k \sim
\lambda^{-1}$, in the non-linear cascade has the form
\begin{equation}
b_k = (\Theta/\kappa^3 \rho)^{1/10} k^{-6/5}, \label{spectrum-pure}
\end{equation}
where $\Theta$ is the flux of energy per unit vortex line length
supported by the non-linear cascade. The value of $\lambda_*$ can
be determined by matching the energy flux $\varepsilon$ with
$\Theta/\rho l_0^2$, where $b_k \sim k^{-1} \sim \lambda_*$. With
Eq.~(\ref{l_0}), we then obtain Eq.~(\ref{lambda_*}).

Kelvin waves decay with emitting phonons \cite{Vinen2000}. For
Kelvin waves of wavenumber $\sim k$, the power of sound emission
per unit line length is given by \cite{SvK_vortex_phonon}
\begin{equation}
\Pi_k \sim \Lambda^6\, \kappa^8\, \rho \; b_k^4 \, k^{11}\, /\,c^5.
\label{sound}
\end{equation}
This dissipation mechanism is negligibly weak all the way down to
wavelengths of order $\lambda_\mathbf{ph}$, given by
Eq.~(\ref{lambda_ph}), where $\Pi_k/\rho l_0^2$ becomes comparable
to $\varepsilon$. The scale $\lambda_\mathbf{ph} \lll \lambda_*$
gives the lower dissipative cutoff of the Kelvin wave cascade.

Now we focus on the strongly turbulent regimes at $r_0 \gtrsim
k^{-1} \gtrsim \lambda_*$. The key quantity here is the energy
transferred to a lower scale after one reconnection of vortex lines
at the scale $k^{-1}$, which, following Ref.~\cite{Sv_95}, can be
written as
\begin{equation}
\epsilon_k \sim f(\gamma) \, \Lambda \; \rho \; \kappa^2 k^{-1} \,
. \label{energy-reconn}
\end{equation}
Here, $f(\gamma)$ is a dimensionless function of the angle
$\gamma$ at which the vortex lines cross,  $\gamma=0$
corresponding to parallel lines. Its asymptotic form is
\begin{equation}
f (\gamma) \sim \gamma^{2}, ~~~~~~~~ \gamma \ll 1 \, .
\label{f_gamma}
\end{equation}

Although, at the scale $r_0$, there is already no coupling between
vortex lines to stabilize the bundles, they should still move
coherently---the geometry of neighboring lines at this scale is
essentially the same over distances $\sim r_0$---until the whole
bundles cross each other. It is possible, however, that vortex
lines within the bundle reconnect. One can show that such
processes can not lead to any significant redistribution of
energy, and thus to a deformation of the bundle at the scale
$r_0$, because they happen at small angles so that the energy
(\ref{energy-reconn}) is too small. Indeed, the dimensional upper
bound on the rate at which two lines at distance $l \ll r_0$ can
cross each other is, from Eq.~(\ref{LIA}), $\Lambda \kappa/r_0 l$,
while the actual value should be much smaller due to the strong
correlations between line geometries. Taking into account that the
number of lines in the bundle is $(r_0/l_0)^2$ and that $\gamma
\sim l/r_0$, the contribution to the energy flux from these
processes is bounded by $(l/r_0)\varepsilon $.

Crossing of the bundles results in reconnections between their
vortex lines and Kelvin waves with somewhat smaller wavelength
$\lambda$ are generated. The coherence of the initial bundles
implies that the waves on different vortex lines must be generated
coherently. Thus, at the scale $\lambda \lesssim r_0$, vortex
lines should be also organized in bundles of length $\lambda$ that
are bent with the amplitude of the Kelvin waves $b_k$, $k \sim
\lambda^{-1}$, while the correlation radius for vortex line
configurations in the transversal direction is $\sim b_k$. Then,
reconnections between the bundles at the scale $\lambda$ transport
the energy to a lower scale. The cascade of bundles should repeat
itself self-similarly in a range of wavelength $l_0 \ll k^{-1},
b_k < r_0$, in which the notion of bundles is meaningful. The
spectrum of Kelvin waves $b_k$ in this regime can be obtained from
the condition $\tilde{\varepsilon}_k \equiv \varepsilon$, where
$\tilde{\varepsilon}_k$ is the energy flux per unit mass
transported by the reconnections at the scale $k^{-1}$ given by
\begin{equation}
\tilde{\varepsilon}_k \sim (k/\rho b_k^2) \; N_k \; \epsilon_k \;
\tau_k^{-1}. \label{flux-k}
\end{equation}
Here, we take into account that the correlation volume is
$b_k^2/k$, the $N_k \sim (b_k/l_0)^2$ is the number of vortex
lines in the bundle, and $\tau_k^{-1} \sim \Lambda \kappa k^2$ is
the rate at which the bundles cross. Physically, $b_k$ determines
the typical crossing angle, $\gamma \sim b_k k$, thereby
controlling the energy lost in one reconnection. Thus, the
spectrum of Kelvin waves in the bundles is
\begin{equation}
b_k \sim r_0^{-1} k^{-2}. \label{spectrum-bundles}
\end{equation}

At the wavelength $\sim \lambda_\mathrm{b}=\Lambda^{1/4} l_0$, the
amplitudes become of order of the interline separation $b_k \sim
l_0$ and the cascade of bundles is cut off. At this scale, $b_k k
\ll 1$, so that the mechanism of self-reconnections is strongly
suppressed. The kinetic times of the purely non-linear regime are
too long to carry the flux $\varepsilon$ \cite{SvK_2004}. We thus
conclude that at $\lambda_\mathrm{c} \lesssim \lambda\lesssim \lambda_b$
the cascade is supported by nearest-neighbor reconnections,  the
amplitudes $b_k$ being defined by the condition of constant energy
flux per unit length and the crossover scale $\lambda_\mathrm{c}$ being
associated with the condition $b_{k\sim 1/\lambda_\mathrm{c}} \sim
\lambda_\mathrm{c}$ meaning that at $\lambda \lesssim \lambda_\mathrm{c}$ the
self-crossing regime takes over. The observation crucial for
understanding the particular mechanism of the cascade and thus
finding $b_k$ is that each nearest-neighbor reconnection
(happening at the rate $\propto \Lambda/\lambda_b^2$ per each line
element of the length $\sim \lambda_b$) performs a sort of {\it
parallel processing} of the cascade for {\it each} of the
wavelength scales of the range $[\lambda_\mathrm{c},\, \lambda_b]$. For the
given wavelength scale $\lambda \sim 1/k$, the energy transferred
by a single collision is $\propto \Lambda (b_k k)^2 \lambda$, and
with the above estimate of the collision rate per the length
$\lambda_b$, this readily yields the estimate $b_k\sim l_0
(\lambda_b k)^{-1/2}$, and, correspondingly $\lambda_\mathrm{c} \sim
l_0/\Lambda^{1/4}$.

In the range $\lambda_\mathrm{c} \gg k^{-1} \gg \lambda_*$, the cascade is
driven by self-reconnections of vortex lines giving the spectrum
$b_k \sim k^{-1}$ \cite{Sv_95}. This regime is replaced by the
purely non-linear regime in the vicinity of $k^{-1} \sim
\lambda_*$ (the actual transition region may be rather wide
\cite{SvK_2004}).

To conclude, the transformation of classical-fluid  Kolmogorov
cascade of superfluid turbulence into the pure Kelvin-wave cascade
requires three intermediate stages associated with locally-induced
motion and reconnections of vortex lines, as illustrated in
Fig.~\ref{f1}.

We thank Victor L'vov and Sergei Nazarenko for drawing our
attention to their work, and are grateful to Nikolay Prokof'ev for
fruitful discussions.


\begin{thebibliography}{99}

\bibitem{Donnelly} R.J. Donnelly, {\it Quantized Vortices in He II}
(Cambridge University Press, Cambridge, 1991).

\bibitem{Cambridge} C.F. Barenghi, R.J. Donnelly, and W.F. Vinen (eds.), {\it Quantized Vortex
Dynamics and Superfluid Turbulence}, Vol. 571 of Lecture Notes in
Physics, Springer-Verlag, Berlin, 2001.

\bibitem{Vinen06} W.F. Vinen, J. Low. Temp. Phys. \textbf{145}, 7
(2006).

\bibitem{Lathrop} G.P. Bewley, D.P. Lathrop, and K.R.
Sreenivasan,  Nature  \textbf{441}, 588 (2006).



\bibitem{3He_Nature} A. P. Finne, T. Araki, R. Blaauwgeers, V. B. Eltsov,
N. B. Kopnin, M. Krusius, L. Skrbek, M. Tsubota, G. E. Volovik,
Nature \textbf{424}, 1022 (2003).


\bibitem{Pickett} D. I. Bradley, D. O. Clubb, S. N. Fisher, A. M. Gue'nault,
R. P. Haley, C. J. Matthews, G. R. Pickett, V. Tsepelin, and K.
Zaki, Phys. Rev. Lett. \textbf{96}, 035301 (2006)

\bibitem{BEC} N. G. Parker and C. S. Adams,
Phys. Rev. Lett. \textbf{95}, 145301 (2005).

\bibitem{Maurer_Tabeling} J. Maurer, P. Tabeling, Europhys. Lett. \textbf{43}, 29
(1998).

\bibitem{Skrbek} S. R. Stalp, L. Skrbek, and R. J. Donnelly, Phys. Rev. Lett.
\textbf{82}, 4831 (1999); L. Skrbek, J. J. Niemela, and R. J.
Donnelly, Phys. Rev. Lett. \textbf{85}, 2973 (2000).

\bibitem{Vinen2000} W.F. Vinen, Phys. Rev. B {\bf 61}, 1410 (2000).


\bibitem{Vinen_Niemela} W. F. Vinen, J. J. Niemela, J. Low Temp. Phys. \textbf{128},
167 (2002).

\bibitem{En_spectra} V.S. L'vov, S.V. Nazarenko, and L. Skrbek, J. Low. Temp. Phys. \textbf{145},
125 (2006).

\bibitem{Nore} C. Nore, M. Abid, and M.E. Brachet, Phys. Rev. Lett. {\bf 78}, 3896 (1997).

\bibitem{Tsubota_2002} T. Araki, M. Tsubota, S.K. Nemirovskii, Phys. Rev. Lett. {\bf 89}, 145301 (2002).

\bibitem{Barenghi_2002} C.F. Barenghi, S. Hulton, and D.C. Samuels, Phys. Rev. Lett. {\bf 89}, 275301 (2002).


\bibitem{SvK_2004} E.V. Kozik and B.V. Svistunov, Phys. Rev. Lett. {\bf 92}, 035301
(2004); \textit{ibid.} {\bf 94}, 025301 (2005).

\bibitem{Vinen_2003} W.F. Vinen, M. Tsubota and A. Mitani, Phys.
Rev. Lett. {\bf 91}, 135301 (2003).


\bibitem{Vinen2001} W.F. Vinen, Phys. Rev. B {\bf 64}, 134520
(2001).

\bibitem{SvK_vortex_phonon} E. Kozik and B. Svistunov, Phys. Rev. B 72, 172505
(2005).

\bibitem{Bottleneck} V.S. L'vov, S.V. Nazarenko, and O. Rudenko, Phys. Rev. B {\bf 76},
024520 (2007).

\bibitem{Sv_95} B.V. Svistunov, Phys. Rev. B {\bf 52}, 3647
(1995).




\end{thebibliography}
\end{document}